\documentclass[preprint,12pt]{aastex}
\usepackage{epsfig}
\usepackage{amssymb,amsmath}

\DeclareMathOperator*{\argmin}{arg\,min}

\iffalse

\else

\fi

\newcommand{\apjvec}[1]{\mbox{\boldmath{$#1$}}}
\newcommand{\apjmat}[1]{{\mathbf{#1}}}

\newcommand{\mI}{\apjmat{I}}
\newcommand{\mA}{{\apjmat{A}}}
\newcommand{\mR}{{\apjmat{R}}}
\newcommand{\Vr}{\apjvec{r}}
\newcommand{\Vc}{\apjvec{c}}
\newcommand{\Vb}{{\apjvec{b}}}
\newcommand{\Vm}{\apjvec{m}}
\newcommand{\Vom}{{\apjvec{\omega}}}
\newcommand{\VomHat}{{\tilde{\Vom}}}
\newcommand{\bi}{{i}}
\newcommand{\wb}{{w_{\bi}}}
\newcommand{\Vcb}{{\Vc_{\bi}}}
\newcommand{\Vrb}{{\Vr_{\bi}}}
\newcommand{\sbs}{{\mathbf{\sigma}^2_{\bi}}}

\begin{document}

\title{Catalog Matching with Astrometric Correction and its\\ Application to the Hubble Legacy Archive}
%\journalinfo{To be submitted to ApJ}

\author{Tam{\'a}s Budav{\'a}ri$^1$ and Stephen H. Lubow$^2$}
\affil{$^1$%Dept.\ of Physics and Astronomy, 
The Johns Hopkins University, 3400 N.~Charles St., Baltimore, MD 21218 \\
$^2$Space Telescope Science Institute, 3700 San Martin Dr., Baltimore, MD 21218}

%\shortauthors{Budav\'ari}
\shorttitle{Catalog Matching with Astrometric Correction}

\begin{abstract}
\noindent
Object cross-identification in multiple observations is often complicated by the uncertainties in their astrometric calibration. 
Due to the lack of standard reference objects,
an image with a small field of view can have significantly larger errors in its absolute positioning than the relative precision of the detected sources within.
We present a new general solution for the relative astrometry that quickly refines the World Coordinate System of overlapping fields.
The efficiency is obtained through the use of infinitesimal 3-D rotations on the celestial sphere, which do not involve trigonometric functions. 
They also enable an analytic solution to an important step in making the astrometric corrections.
In cases with many overlapping images, the correct identification of detections that match together
across different images is difficult to determine. 
We describe a new greedy Bayesian approach for selecting the best object matches across a large number of overlapping images. The methods are developed and demonstrated on the Hubble Legacy Archive, one of the most challenging data sets today.
We describe a novel catalog compiled from many Hubble Space Telescope observations, where
the detections are combined into a searchable collection of matches that link the individual  detections. The matches provide descriptions of astronomical objects involving multiple wavelengths and epochs. 
High relative positional accuracy of objects is achieved across the Hubble images, often sub-pixel precision in the order of just a few  milli-arcseconds. The result is a reliable set of high-quality associations that are publicly available online. 
\end{abstract}


\email{budavari@jhu.edu, lubow@stsci.edu}

\section{Introduction} \label{sec:intro} 
\noindent
Astronomical research is benefiting from  the use of large searchable
catalogs made possible by advances in detector and computer technology. 
With the rapidly growing capacity of CCDs,  astronomical objects are being
detected at an increasingly fast rate. A major challenge to astronomy is
being able to synthesize this large amount
of data into an organized set of information that takes the form of a catalog.
The catalog describes
not only the properties of the images, but more importantly the sources detected within them. 
In its simplest usage, one can quickly determine whether an object
is found in some region of space and examine its listed properties. But there are many
other possible uses of a large catalog. For example, non-positional (all sky) searches
for objects with particular properties can provide statistical information
on large numbers of objects. Objects that lie at the extremes of the distributions provide
the potential for new discoveries. 

Advances in relational database technology
have provided a means for building, storing, and searching large catalogs. 
This technology alone, however, is not sufficient. 
One reason is that the rate at which data is becoming available from detectors is on
a path to exceed storage and processing capabilities of computers for a large class of problems \citep{wwt01}.
New approaches to data organization and new algorithms are needed to deal
with the challenges of a telescope such as Hubble Space Telescope (HST).  
In the case of the HST, 
some of these arise from 
its small field of view and 
the complex
geometry of its overlapping exposures.
%
%HST has taken
%many observations that form a sparse, complex geometric pattern on the sky of small, partially overlapping exposures.
%The overlapping exposures were taken over irregular time intervals and involve different detectors, filters, orientations,  
%and exposure times. 
%
%In this paper we describe new approaches along these lines
%and their application to the construction of a catalog for the Hubble Space Telescope (HST). 
 
Crossmatching of detected sources from images taken at different wavelengths 
and at different epochs is a crucial capability for catalog construction. By matching
source detections across multiple images to a single astronomical object, one can then determine spectral
and temporal properties of the object. 
The statistical methodology introduced by \citet{pxid} provides a clean framework for determining symmetric $n$-way associations, and has been successfully applied in several studies including \citet{heinis}, \citet{roseboom}, \citet{kerekes}, \citet{rots} and \citet{budavari_sn}.
Performing crossmatching on HST observations also involves
adjusting the positions of the images to place them into better alignment. 
%It also involves determining which sources match together. 
In practice, these two are related, since the accuracy of the alignment is determined
by how well the sources match together.

One approach to crossmatching involves registering images against a known
catalog, such as Guide Star Catalog (GSC II; \citealt{Lasker08}) or the Sloan Digital Sky Survey \citep[SDSS;][]{York00} 
SkyServer database.\footnote{Visit the SkyServer online at \url{http://skyserver.sdss.org/}}
In this approach, the absolute position information
in the catalog is used to anchor positions of matching sources detected in the images.
Sources from different images can then be compared. 
The drawback to this method is that few or none of the sources in the catalog
may be in the image. This situation is particularly true of images
taken by the HST which has a small field of view.
On the other hand, it often detects many more sources than have been previously detected.
A different approach, the one we adopt here, is to crossmatch the many sources in overlapping images taken by
HST against each other, rather than against an existing catalog. 
The process involves adjusting the positions of the overlapping images to improve the residual
errors in the crossmatching.
In this case, relative, rather than absolute, astrometry is determined
involving many detected sources.  Absolute astrometric positioning can then
later be determined by matching the set of overlapping images
as a larger unit against the absolute standards.

Some astronomical projects, such the SDSS, were designed with the goal of providing a catalog.
The observations are made in a way that uniformly tiles regions of sky in certain filter
bands and at certain regular time intervals. On the other hand, the HST
as well as other major space observatories, have generally targeted
particular sources, although particular programs have undertaken surveys
for very limited regions of the sky. 
For more than two decades,
images have been taken for many
independent programs, resulting in a sparse, complex geometry of sky coverage and
in irregular time intervals between observations of objects. 
In many cases, the coverage involves
partially overlapping exposures with a variety of outline sizes
and shapes, orientations, filters, and exposure times.
The HST has provided some of the highest resolution images ever obtained. Therefore,
in spite of the challenges, there is a potentially important scientific gain
by having a catalog for HST. 

The Hubble Legacy Archive\footnote{Visit the Hubble Legacy Archive at \url{http://hla.stsci.edu/}}
\citep[hereafter HLA; ][]{Jenkner06}
provides enhanced data products and advanced browsing capabilities online. Its products include lists of detected sources and their photometric properties\footnote{See source list description at \url{http://hla.stsci.edu/hla\_faq.html\#Source1}} \citep{Whitmore08, Lindsay10}.
These source lists are obtained running DAOPhot  \citep{Stetson87}  and Source Extractor  \citep{Bertin96}
software on combined, drizzled images.
Each of these images is the result of combining exposures for a single instrument, detector and filter
from a single visit (pointing of HST). 
These source lists contain information about source positions, fluxes, magnitudes, morphology, etc.
The source lists and auxiliary HLA data provide the needed input information to implement our algorithms for the Hubble catalog.
By crossmatching the source lists based on these images, we obtain multi-wavelength time-domain
information about astronomical objects.

In this paper we describe some novel approaches to crossmatching with application to the construction of a catalog for HST.
In Section~\ref{sec:xmatch} we describe general algorithms
for image clustering, astrometric correction, and source aggregation into matches.
Section~\ref{sec:catalog} describes the pipeline that we have constructed to build an HST catalog 
based on these algorithms. Section~\ref{sec:results} contains an analysis of the properties
of this catalog for ACS/WFC and WFPC2. Section~\ref{sec:summary} concludes our study.

\section{Improved Matching and Astrometry} \label{sec:xmatch} 
\noindent
To create a reliable set of associations, we need high-precision astrometry. Solutions exist for the global World-Coordinate System \citep[WCS;][]{wcs} astrometric determination of large images \citep[e.g.,][]{Hogg08}.
However, small images are still difficult to work with, especially when the sources are very faint, since they typically do not contain a sufficient number of calibration standard objects. The only possibility is to cross-calibrate the set of small overlapping images to obtain improved relative astrometry. Once several of the small images are locked in and tied together, the number of available standards will increase that enables a more accurate absolute positioning. First we discuss how to cross-calibrate two or more images to improve their relative accuracy. Later in the section, we describe a Bayesian method  that determines the matching sources  from sets of many nearby detections in overlapping images.

\subsection{Finding Pairs} \label{sec:pairs} 
\noindent
Crossmatching millions of sources contained in many irregularly placed images on the sky is a computationally challenging task.
A substantial reduction in computational overhead is obtained by determining disjoint sets of overlapping images through a friends-of-friends (FoF) algorithm.
We use the term {\it mosaic} to describe each of these disjoint sets. Since the mosaics are disjoint, the source matching is carried out independently within each and every one of them. 
The first step in creating sets of matched sources involves the
identification of pairs of sources  that reside in different images and are close together on the celestial sphere.
A tolerance is then needed to be applied that depends on the accuracy of the relative astrometry.

\subsection{Astrometric Corrections} \label{sec:astrometry} 
\noindent
Given the set of matched pairs in a mosaic, determined with some tolerance in separation,
we next adjust the relative astrometry of the images that make up the mosaic, in order
to reduce the separations of sources  in the pairs.
The traditional approach is to apply corrections to the World-Coordinate System, which is often very expensive computationally, and involves many trigonometric function evaluations. Here we choose a different approach, which is faster to calculate, and can accomplish multiple objectives in just one step. 
It is common practice to consider translations and rotations of the images in their X-Y plane. 
Our approach, however, is to use three-dimensional rotations on the celestial sphere. Such transformations can account for both rotation and translation locally in the tangent plane. When the axis of the 3-D rotation is parallel with the pointing of a given image, the transformation indeed corresponds to a 2-D rotation. But if the axis is perpendicular to the pointing, the 3-D rotation results in a translation in the tangent plane. If we allow for any direction, a single transformation can describe a combined effect. Here we work with 3-D normal vectors, which is often the preferred way in spherical calculations.

We first consider an idealized problem where a single image is rotated in 3-D to minimize the separations of
its sources 
from those matched to a fixed reference image.
Let $\Vr_{\bi}$ represent the direction of the $\bi$th detection in the image, and let $\Vc_{\bi}$ be a matching  reference direction. We can form a set of $(\Vr_{\bi},\Vc_{\bi})$ pairs to be used in the astrometric correction. 
Now we have to solve an optimization problem for the 3-D rotation $\mR$. The transformed position would be $\Vr' = \mR\,\Vr$. Hence the optimization formally is
\begin{equation}
\tilde{\mR} = \argmin_{\mR \in {\rm{}SO(3)}} 
\left\{ %\frac{1}{2} 
	\sum_{\bi} \wb \Big[\Vcb - \mR\,\Vrb \Big]^2 
\right\}
\label{mR}
\end{equation}
where $\wb$ is the precision parameter that is related to the $\sbs$ accuracy via $\wb=1/\sbs$. 
Much like the potential energy stored in springs, our cost is quadratic in the displacements, and can be thought of as a system of springs, where the spring constants are proportional to the $\wb$ weights. The solution is the equilibrium position of the sphere with a fixed center, which is given by the $\tilde{\mR}$ rotation matrix.

This general optimization, however, is computationally expensive. 
For this reason, working with orthonormal matrices is not practical. But for small corrections, {\em{}infinitesimal rotations} are ideally suited. 
Infinitesimal rotations have many advantageous mathematical properties over general 3-D rotations. For example, they are commutative. 
Let $\Vom$ represent the axis and the angle of the rotation. The axis is defined by the direction and the angle is the length of the vector. The infinitesimal transformation is then
\begin{equation}
\Vr' = \Vr + \Vom\times\Vr.
\end{equation}
Considering that the change \mbox{$\Delta\Vr=\Vom\times\Vr$} is perpendicular to $\Vr$ and small in amplitude, the resulting vector $\Vr'$ stays approximately normal to quadratic order in $\Delta r$.
The optimization problem with the infinitesimal rotation becomes analytically tractable. 
The cost function, whose minimization yields the $\VomHat$ estimate, is now not only quadratic in the displacement but also in its $\Vom$ parameter,
\begin{equation}
\VomHat = \argmin_{\Vom \in \mathbb{R}^3} 
\left\{ %\frac{1}{2} 
	\sum_{\bi} \wb \Big[\Vcb - \big(\Vrb + \Vom\times\Vrb\big) \Big]^2 
\right\}.
\label{vomhat}
\end{equation}
By requiring that the gradients equal zero, we obtain a 3-D linear equation
\begin{equation}
\mA \VomHat = \Vb
\label{omeq}
\end{equation}
which is a result of the summations
\begin{eqnarray}
%\mA & = & \sum_{\bi} \wb\,\big( \Vrb^2\,\mI - \Vrb \otimes \Vrb \big) \\
\mA & = & \sum_{\bi} \wb\,\big( \mI - \Vrb \otimes \Vrb \big) \\
\Vb & = & \sum_{\bi} \wb\,\big( \Vrb \times \Vcb \big)
\end{eqnarray}
where $\mI$ is the identity matrix, and $\otimes$ is the operator of the dyadic product.
A derivation of Equation~(\ref{omeq}) is given in Appendix~\ref{app1}.

In keeping with the spring analogy described below Equation~(\ref{mR}),
Equation~(\ref{omeq}) can be interpreted as a torque equation.
Vector $\VomHat$ can be considered an angular acceleration vector of a sphere 
at an early time after its release that
is related to its angular displacement at this early time.
The sphere that has springs connecting each point $i$ of mass $\wb$ located at position $\Vr_{\bi}$ to
the fixed standard located at $\Vc_{\bi}$. 
 Matrix $\mA$ is the moment of inertial tensor for
sources that lie on the unit sphere that is subject to a torque given by  $\Vb$.
Notice that in the equation for
$\Vb$, quantity $\wb \Vc_{\bi}$ can be replaced  by $\wb (\Vc_{\bi}- \Vr_{\bi})$, which is the force due to the spring.

Since $\mA$ is a \mbox{$3\!\!\times\!\!3$} matrix,
Equation~(\ref{omeq}) has a simple analytic solution for $\VomHat$, e.g., by applying
of Cramer's rule for matrix inversion. 
In practice, the matrix inversion can be performed numerically by one of several possible methods. 
The cost of the aggregation is linear in the number of pairs, and the matrix inversion involves a small number of operations. 
Therefore, the computational overhead for this approach is likely to be close to optimal.

So far we assumed that there is a set of $\{\Vcb\}$ calibrators in a fixed, perfect reference image, which is clearly not true in our case. Instead we have 
imprecise source positions contained in groups of overlapping observations, the mosaics. 
In each mosaic we want to correct every one of the images,  but the combined minimization problem is not as simple as it is for a single image. 
Since we are only concerned with relative precision, different heuristic approaches come to mind to work around the lack of a true set of reference positions. For example, one of the images could be singled out to act as a reference frame onto which all others are corrected. In a group of images, however, there is no guarantee that there is a single image with which all others would overlap, because a mosaic consists of friends of friends. Another option would be to use the average positions of a preliminary match. We follow a third approach where the correction of a given image is based on all the other images.
One-by-one we consider each image and derive their corrections independently, effectively assuming that all the others are perfect. Hence we use them as calibrators. The matched pairs being applied to Equation (\ref{omeq}) are then all
the pairs
for the mosaic involving the image being corrected.  Having derived the infinitesimal rotation for all images, we update the source positions with the appropriate corrections and repeat until convergence. The $\Vom$ vectors are accumulated, i.e., summed up, for each image (cf.\ commutativity of infinitesimal rotations) and saved for future reference. This also enables us to safely work with clean samples of stars, but apply the correction for all sources afterwards. One can also use this information to correct the alignment of the images in their WCS headers to reflect the changes.

%The convergence of the procedure is fast, but can be affected by erroneous initial matches. To avoid such situations, we exclude at run time all the pairs whose angular separation grows too large during the iterative optimization, as opposed to shrinking as expected.

\subsection{Friends-of-Friends Source Aggregation} \label{sec:fof} 
\noindent
Now we use the astrometrically improved coordinates to perform a final cross-identification. This time we apply a smaller tolerance than the initial crossmatch, and again produce matched pairs of sources, as discussed in Section~\ref{sec:pairs}.
Once all close source pairs are identified across images, the next step is to run a friends-of-friends (FoF) algorithm, 
also known to statisticians as single-linkage hierarchical clustering,
to link all nearby detections into singly connected graphs. 
Such clustering is very efficient and can be easily implemented in any programming environment.

The FoF groups enumerate all the sources that can be linked together using a specified pairwise distance threshold. There is no guarantee, however, that this algorithm  provides statistically meaningful matches. For example, by linking nearby sources along a line, we can potentially construct very long chains whose ends are far away from each other 
\citep{everitt2011book}. These FoF matches are just considered preliminary and need to be studied further.

We note that other hierarchical clustering methods also exist 
\citep{everitt2011book},
whose performance depends on the nature of the data and the goals of
the project. Here we look for clusters only to provide candidates for  
the cross-identification.
Preliminary experiments with several off-the-shelf tools, including average and complete linkage,
were performed on select areas without noteworthy accuracy or speed
improvements. These other tools also provide clusters with a wide range of
sizes and shapes.
For cross-matching, we are interested only in identifying isolated compact clusters. 
To accomplish this, we apply
a Bayesian method to evaluate the likelihood of various possible
clusters based on the separation of the members and the estimated
positional errors.

\subsection{Comparing Configurations} \label{sec:comparison} 
\noindent
The question is whether breaking up the FoF groups into smaller ones could yield better models, and if the answer is yes, which configuration would the best. 
We can address the issue by applying a Bayesian model comparison. If $D$
represents the entire set of $n$ positions in a given FoF group, we can consider alternative hypotheses, where $D$ is partitioned into disjoint components that correspond to separate objects.
The baseline case is when we have a single component. But there could be as many components as the number of detections. 
In general, let $G$ represent a hypothesis with $\Gamma$ components with model directions $\{\Vm_{\gamma}\}$ for \mbox{$\gamma\!=\!1...\Gamma$}.
Also let $S_{\gamma}$ be the list of sources that are assigned to each component and $D_{\gamma}$ their measured directions such that
\begin{equation}
D = \left\{\Vr_1,\Vr_2,\dots,\Vr_n\right\} = \bigcup_{\gamma=1}^{\Gamma} D_{\gamma}
\end{equation}
and
\begin{equation}
D_{\gamma_1} \cap D_{\gamma_2} = \emptyset \ \ \ \ {\rm{}when} \ \ \ \gamma_1\neq\gamma_2.
\end{equation}
%\begin{equation}
%p(D|G) = \prod_{\gamma=1}^{\Gamma}\,\int\,p(m_{\gamma}|H)\,p(D_{\gamma}|m_{\gamma},G)\ {\rm{}d}{}m_{\gamma}
%\end{equation}
% 

Similar to the problem of probabilistic cross-identi\-fi\-ca\-tion discussed by \citet{pxid}, the likelihood of $G$ can be derived from the astrometric uncertainties and the prior on the directions of the objects.
In general, it is essentially just the product of the individual $\Gamma$ matches
\begin{equation}
p(D|G) = \prod_{\gamma=1}^{\Gamma}\,\int{}p(\Vm_{\gamma}|G)\!\prod_{\bi\in{}S_{\gamma}}\!p_i(\Vr_{\bi}|\Vm_{\gamma},G)\ {\rm{}d}{}\Vm_{\gamma}.
\end{equation}
The calculation is analytic if the uncertainty is modeled with the spherical normal distribution 
\citep{fisher53} 
\begin{equation}
p_i(\Vr_i|\Vm_{\gamma},G)  = \frac{w_i\,\delta(|\Vr_i|\!-\!1)}{4\pi \sinh w_i}\,\exp \big( {w_i\Vr_i  \cdot \Vm_{\gamma}} \big)
\end{equation}
and the prior is isotropic
\begin{equation}
p(\Vm_{\gamma}|G) = \frac{1}{4\pi}\,\delta(|\Vm_{\gamma}|\!-\!1),
\end{equation}
where $\delta(\cdot)$ is the Dirac $\delta$-function.

We can derive the improvement over the case $S$ when all detections are considered separate, cf.~\citet{pxid}.
In the limit of high astrometric accuracies, \mbox{$w_i$$\gg$1}, the Bayes factor \mbox{$p(D|G)/p(D|S)$} becomes
\begin{equation}
%\frac{p(D|G)}{p(D|S)} 
B = 2^{n-\Gamma}\!\!\left(\prod_{i=1}^n{}w_i %\frac{\delta(r_i -1)}{4 \pi}
\!\right)\!
\prod_{\gamma=1}^{\Gamma}\frac{\exp \left(\frac{\sum w_i w_j \psi_{ij}^2}{-4\sum{w_i}} \right)}{\sum{} w_i} 
%p(D|G) = 2^{n-\Gamma}\!\!\left(\prod_{i=1}^n{}w_i\!\right)\!%
%\times \nonumber \\ & &\times  
%\prod_{\gamma=1}^{\Gamma}\frac{\exp \left(\frac{\sum w_i w_j \psi_{ij}^2}{-4\sum{w_i}} \right)}{\sum{}w_i} ,
\label{eq:bf1}
\end{equation}
where $\psi_{ij}$ is the angular separation between the $\Vr_i$ and $\Vr_j$ vectors,
and the unmarked $i$ and $j$ summations are over the corresponding $S_{\gamma}$ sets. Note that this equation is also valid for singleton groups with only one detection because the summation includes the $i$=$j$ cases, for which the separation is 0 by definition.

Other subdivisions of an FoF group can be evaluated against each other using the corresponding Bayes factors, the ratios of their likelihoods, which can be directly evaluated from the previous formula. For example, comparing $G_1$ and $G_2$ yields
\begin{equation}
B_{12} = \frac{p(D|G_1)}{p(D|G_2)}.
\label{eq:bf2}
\end{equation}
where the $p(D|S)$ baseline case simply cancels out.
When the value is 1, the case is indecisive but if \mbox{$B_{12}\!>\!1$}, the measurements prefer the $G_1$ hypothesis; otherwise $G_2$ is favored.

While the statistical approach for selecting the best components is clear in principle, the problem is still impossible to fully solve in practice due to its combinatorial nature. For FoF associations of, say, only 20 sources, the computational cost of exploring all possible combinations is simply prohibitively large.
Thus we resort to a greedy algorithm that essentially corresponds to a shrinking pairwise distance threshold.
Each FoF group is represented as a connection graph, where the edges link the nearby sources. We build the initial graph of the group, and evaluate its $\Gamma$=1 baseline likelihood. 
Next we repeatedly break the longest edge until the graph splits.
We then evaluate the new hypothesis of these two components ($\Gamma$=2) against the baseline.
If the new model is more likely, i.e., the Bayes factor is greater than unity, we save it along with (the logarithm of) its likelihood. 
Afterwards we repeat the same steps to further break the subgraphs until we end up with separate objects with no connections at all. 
The result is a set of possible associations that are better than the original. Out of these we can pick the most likely.

\section{Pipeline for the Hubble Catalog} \label{sec:catalog} 
\noindent
The HLA contains metadata extracted from HST images. They are stored in a commercial Microsoft SQL Server 2008 database. The overall design builds on the success of the Sloan Digital Sky Survey's SkyServer archive
%\footnote{Visit the SkyServer online at \url{http://skyserver.sdss.org/}}
and applies a number of its extensions developed for spatial searches and spherical geometry. In this section, we describe our implementation of the aforementioned algorithms that runs completely inside the database engine, which in turn provides efficient parallel execution and rapid access to derived data products.

The HST pointings generally do not follow any particular tiling pattern on the sky. During its long life, the Hubble Telescope has taken over a half-million exposures  date, and many of these overlap with others from different programs. We can determine the overlapping exposures based on their intersecting sky coverage. The geometry of all visits is stored inside the database using the Spherical Library, along with its SQL routines and spatial indexing facilities \citep{spherical}.
%
%We build mosaics from visits that overlap. The spatial searches are sped up by the Hierarchical Triangular Mesh \citep[sHTM;][]{htm, spherical}. 
%
%We will refer to these groups of overlapping images as {\em{}mosaics}. 
%To clarify, we do not actually build mosaic images but instead define a mosaic as a maximal set images that overlap. 
%

The HLA contains source lists that are based on applying combined drizzled image files to 
DAOPhot and  Source Extractor software. They provide positional and other information for each
detected source.
We have carried out the crossmatching on the set of Source Extractor source lists available from DR6 of the HLA
for  ACS/WFC and WFPC2 taken together.   %
Furthermore, we only consider sources that are determined to be good quality by Source Extractor, which means that the {\tt{}Flags}%
\footnote{Description of {\tt{}Flags} is at \url{http://hla.stsci.edu/hla\_faq.html}}
value is less than 5. The crossmatching process operates on each set of source lists in a mosaic.  

The absolute astrometry of the HLA images is determined by the pointing
of accuracy of HST, which is about 1 to 2 arc-sec in each coordinate.
The HLA attempts to correct the absolute astrometry%
\footnote{Description of HLA astrometry is at \url{http://hla.stsci.edu/hla\_faq.html\#astrometry}}
further by
matching against three catalogs: Guide Star Catalog 2 (GSC2), 2MASS,
and the Sloan Digital Sky Survey (SDSS) catalog.  The typical absolute
astrometric accuracy of the HLA-produced images is $\sim$0.3 arcsecs in each
coordinate.
The majority of images, about 80\% of ACS/WFC, are corrected in this
way.

\subsection{White-Light Sources and Images} \label{sec:whl}
\noindent
The source detections made in the HLA are obtained in two steps. 
First, all images within a visit (HST pointing) for a given detector 
are combined to produce a ``white-light'' detection image, the deepest image
possible for the visit. 
Source Extractor is then run on the detection image to identify source positions. In the second step,
Source Extractor is run on each ``color'' (single filter) combined image within the visit for each detector to find sources at the positions identified by the detection image. If a source is 
found at an indicated position, then its properties are determined and stored in a database table. 
There are currently about 49,000 source lists containing about 46 million source detections. 

Given the above procedure for extracting sources, it makes no sense to crossmatch them
in different filters for the same detector and visit. The reason is that, for a given detector and visit, color sources are constructed to have
exactly the same position when they exist in different filters. Therefore, crossmatching must
be carried out across visits and detectors. 
To do this, for each detector, we need to use visit level white-light source lists
and the geometric outlines for the visit level combined images. 
The geometric outlines are used to determine the mosaics described in Section~\ref{sec:pairs}.
Both the white light source lists and the geometric outlines
 are determined from available source lists and image metadata that reside in the HLA database.

\begin{figure}
\epsscale{0.5}
\plotone{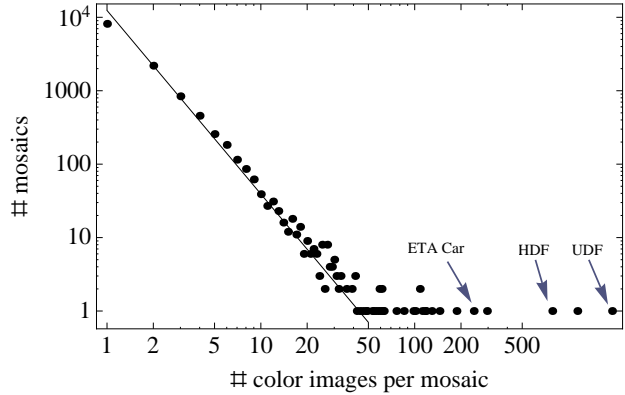}
\caption{Log-log plot of the number of mosaics versus the number
of  white-light (combined-filter visit-level ACS/WFC and WFPC2) 
images per mosaic. The line follows a power law with index -2.5. 
Some popular HST programs are indicated and lie in the long tail of the distribution. Crossmatching within these cases provides a computational challenge.}
\label{fig:mossize}
\end{figure}

Fig.~\ref{fig:mossize} plots the distribution of the number of white-light images within a mosaic. For smaller mosaics, the results follow a power law decline in frequency with number of images. The fall-off occurs with an index of about -2.5 and approaches unity at around 50 images. There is, however, a long tail that extends beyond that to almost 2000 images.
This tail involves many of the large HST programs, such as the Ultra Deep Field (UDF).
Much of the computational challenge lies in the crossmatching within the mosaics of this tail.

\subsection{Mosaics and Source Pairs} \label{sec:pipepairs}
\noindent
Mosaics are constructed by the application of the algorithms described in Section \ref{sec:pairs}.
The determination of mosaics is efficiently accomplished by using the Hierarchical Triangular Mesh indexing
\citep[HTM;][]{htm} to determine white-light images that are close to a given image 
and by using the Spherical Library \citep{spherical} to determine whether the images actually overlap. 
The pipeline completes the cross-matching on mosaics sequentially. That is, the crossmatching
of sources is carried out for one mosaic at a time.
Within each mosaic, the pipeline determines pairs of white-light sources whose separations are less than 0.3 arc-sec. 
This threshold is required to accommodate some of the more significant systematic offsets between the images. Larger offsets are currently not considered in this study.
This matching can be carried out efficiently in an SQL database that is indexed on RA and Dec.

\subsection{Refined Astrometry} \label{sec:ftrans}
\noindent 
Using the white-light source pairs, the pipeline carries out astrometric corrections of the images within a mosaic using
the algorithms  described in Section  \ref{sec:astrometry}. 
The astrometric correction algorithm that uses infinitesimal 3-D rotations is implemented in the C\# programming language and is integrated directly into the database. The query that selects sources for astrometric correction is limited to detections that are likely stars based on the morphology. Despite the risk presented by the resolved objects, we also experimented with using the both resolved and unresolved sources also, which has the advantage of having more sources. But we obtained similar global results in most cases. The details of the sample selection for the correction might change in the future. 
As described in Section \ref{sec:astrometry}, 
the astrometric correction procedure iterates through all images in a mosaic and corrects them one-by-one assuming the others are correct. The initial large angular separation limit is decreased over time. In this way, any possible wrong associations (due to the generous threshold) are later eliminated 
as the relative calibration of the images tightens up.
The convergence of the procedure is fast, but can be affected by erroneous initial matches. To avoid such situations, we exclude at run time all the pairs whose angular separation grows too large during the iterative optimization, as opposed to shrinking as expected.
The transformations are persisted in a database table and are available to future pixel-level corrections that can potentially improve the quality of a combined image.
Using the refined astrometry of all images, the cross-identification is performed once more with similarly large thresholds, so the list of matching pairs of sources is as inclusive as possible.

\subsection{Best Matches} \label{sec:bestmatch}
\noindent
The pairwise matches are quickly grouped into FoF clusters that are obtained by using the sorted tables with appropriate primary keys.
These clusters then provide preliminary sets of matched sources, which might
contain large chains that may then be split by the Bayesian method described
in Section \ref{sec:comparison}. 
This probabilistic splitting is carried out by an SQL stored procedure called the {\em{}chainbreaker}, which is also
written in the C\# language. It reads all the sources grouped into matches and builds a graph out of them. Using the refined astrometric positions, we can safely ignore large separations and start with a 0.1 arc-sec linking length. For each match, the splitting procedure considers a number of cases, using the greedy algorithm of Section~\ref{sec:comparison},  all the way to separate objects.
Each result that proves to be better than the unsplit case is saved in a database table.

For each match, the {\em{}chainbreaker} selects the most favorable case based on equations~(\ref{eq:bf1}) and (\ref{eq:bf2}), and stores the results in database tables that are used for catalog construction. The output of the stored procedure includes all the values that are required to link to the HLA database and to fetch any relevant details of the observations related to the sources.

The astrometric correction, FoF and chainbreaker steps are conceptually intertwined. After completion of the entire procedure, one could analyze the astrometric uncertainties and revise the results by running the procedure repeatedly. This is a possible direction of further studies.
A completely integrated scheme, where the chainbreaker is applied in each astrometric iteration, is impractical due to its computational cost.

\subsection{Catalog}
\noindent
After the matching of the white-light sources is completed,
we apply the available color (filter) level information for each white-light source and store the results in a database table for the catalog.
The source catalog we construct then contains the information about the source detections in each filter of a visit.
We also include in the catalog the cases of color non-detections, i.e., color dropouts, within a visit.
Such non-detections can be inferred from the two-step source detection process described above, since
we know which images were used to build the white-light detection image and which images
contained a source detection at a given position.

Consider the case that a user specifies
a region-based (e.g., position and search radius) search of the catalog.
Another form of non-detection arises in this
case when no source is found within a white-light image that overlaps with this user specified region.
All such white-light non-detections cannot
be stored in a database, but instead are inferred at run time within the catalog search function we have
developed. We therefore provide information about detections and two types
of non-detections (color and white-light).

The matches are a collection of color source detections and color non-detections (as described above) that lie
sufficiently close together in the sky. (White-light nondetections do not belong to a match.)
Each match consists of at least one detected source. Each match can be considered to
be describing a single physical object whose properties are determined by the properties of sources
in the match. For example, each match has a certain position associated with it (MatchRA, MatchDec) that is determined by its source positions. This position is considered to be the location of the object described by the matched sources.

Not all matches involve different visits or detectors. For example, images that are crossmatched
generally contain some sources do not match those in an image obtained from a different detector or visit.
For other cases, the source could not be crossmatched because the image containing the source did not overlap sufficiently with an image from another visit or detector. About 60\% of the images with source lists
could not be crossmatched with other images. The sources in these images are also
included in the catalog for completeness.  In this case, the matches only involve a single detector and visit.

The crossmatching of the HLA DR6 ACS/WFC and WFPC2 source lists was carried out with SQL Server 2008 running on a single (aging) Dell PowerEdge 2950 server with Intel Xeon E5430 @ 2.66Ghz (2 processors) and 24GB of memory.
The entire pipeline for matching the sources and building the catalog runs automatically, and takes a few days to complete.

\section{Results} \label{sec:results} 
\begin{figure}
\epsscale{1}
\plottwo{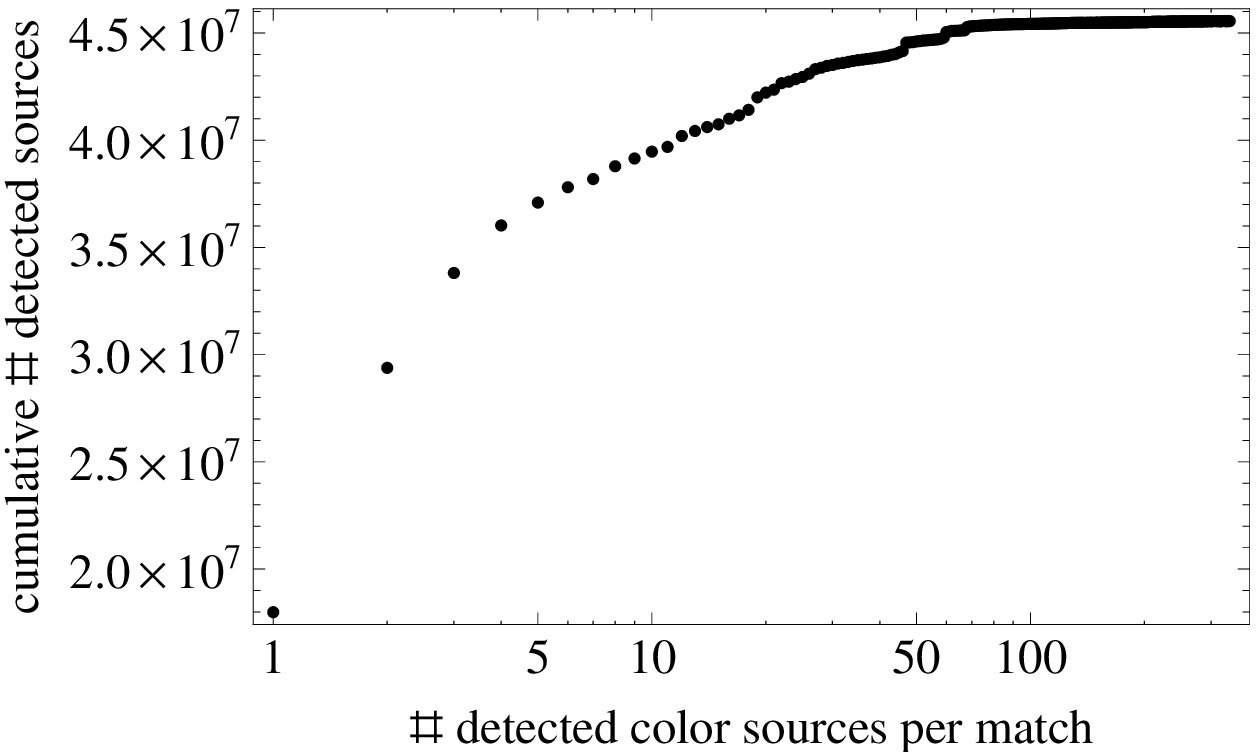}{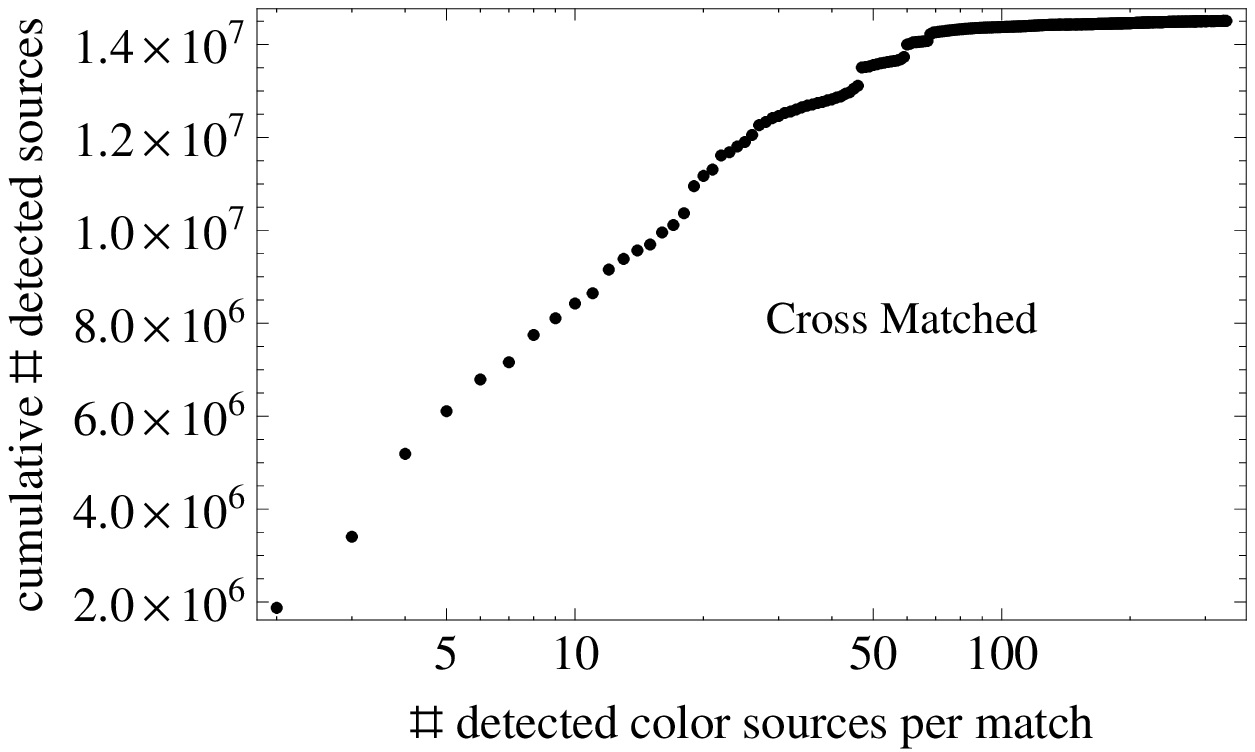}
\caption{Left: Semi-log plot of the cumulative number of detected  color sources as a function of the
number of detected color sources in a match. Right: As in left panel, but only for sources that are crossmatched
with sources in other visits or detectors. }
\label{fig:matchsize}
\end{figure}

\noindent
The final catalog is comprised of entries for over 45 million sources detections and 15 million color dropouts for ACS/WFC and WFPC2.
The catalog indicates which sources match together.
Among the  matches that involve more than one visit, there are on average 5.4 detected (color) sources,
which on average involve 3.8 different visits. 
The left panel of Fig.~\ref{fig:matchsize} plots the cumulative number of detected sources as a function of the
number of detected sources in a match.
About 40\% of the sources are in matches with only one detected  source. About 30\% of these
cases involve
non-overlapping or insufficiently overlapping images. About 20\% of the sources are in matches with 5 or more detected sources.
About 2\% of the sources lie in matches with 50 or more sources. 
The right panel of Fig.~\ref{fig:matchsize}  is similar to the left panel, but restricted to crossmatched 
sources. In this case, there are no matches with
less than 2 sources, as required for crossmatching. All these sources are in images
that have been astrometrically corrected by our algorithm. About 6 million cross-identified detections lie in matches with more than 10 sources.

\begin{figure}
\epsscale{0.5}
\plotone{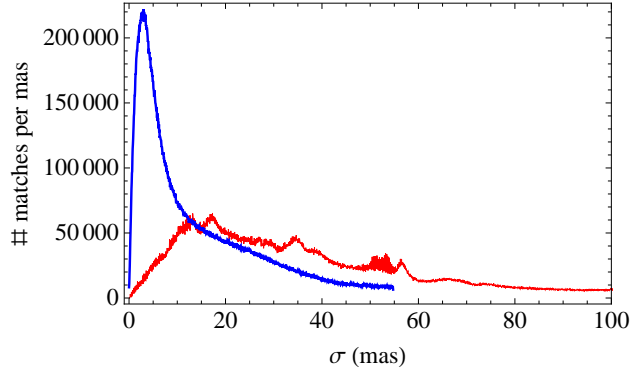}
\caption{Plot of the distribution of the standard deviations of the white light source positions per match
for matches with more than one white light source.
The upper (lower) curve
is after (before) astrometric correction. The lower curve is the current relative astrometry provided by HST images. }
\label{fig:sigmaC}
\end{figure}

For each set of matching sources, we determine the standard deviation of the positions among the
white light sources involved in the match, where more than one white light source is involved.
(The white light sources are used because they are involved in the crossmatching and are statistically independent, unlike the color sources.)
The distribution of the standard deviations for all matches involving more than one white-light source
is shown in Fig.~\ref{fig:sigmaC}. The lower curve is based
on the the current HST astrometry for these matching sources, 
i.e., positions before our astrometric correction is applied.
The upper curve is based on positions after astrometric correction. 
The median before astrometric correction is 32.5 mas.
The peak (or mode) of the astrometrically
corrected distribution is 3.0 mas, and the median is 9.1 mas.

In order to test the validity of these matches, we compare the values of the fluxes
determined by Source Extractor within radii of 3 pixels of the source centers (FluxAper2)
for pairs of source detections in the same match that have the same instrument, detector and filter.
No constraints are applied on the exposure times of the sources in the pairs.
Apart from variable objects that are rare, we expect the flux difference to be small
if the match corresponds to a single object. Flux differences may also be caused by detector degradation over time. We do not attempt to account for this effect in this paper.

\begin{figure}
\epsscale{1}
\plottwo{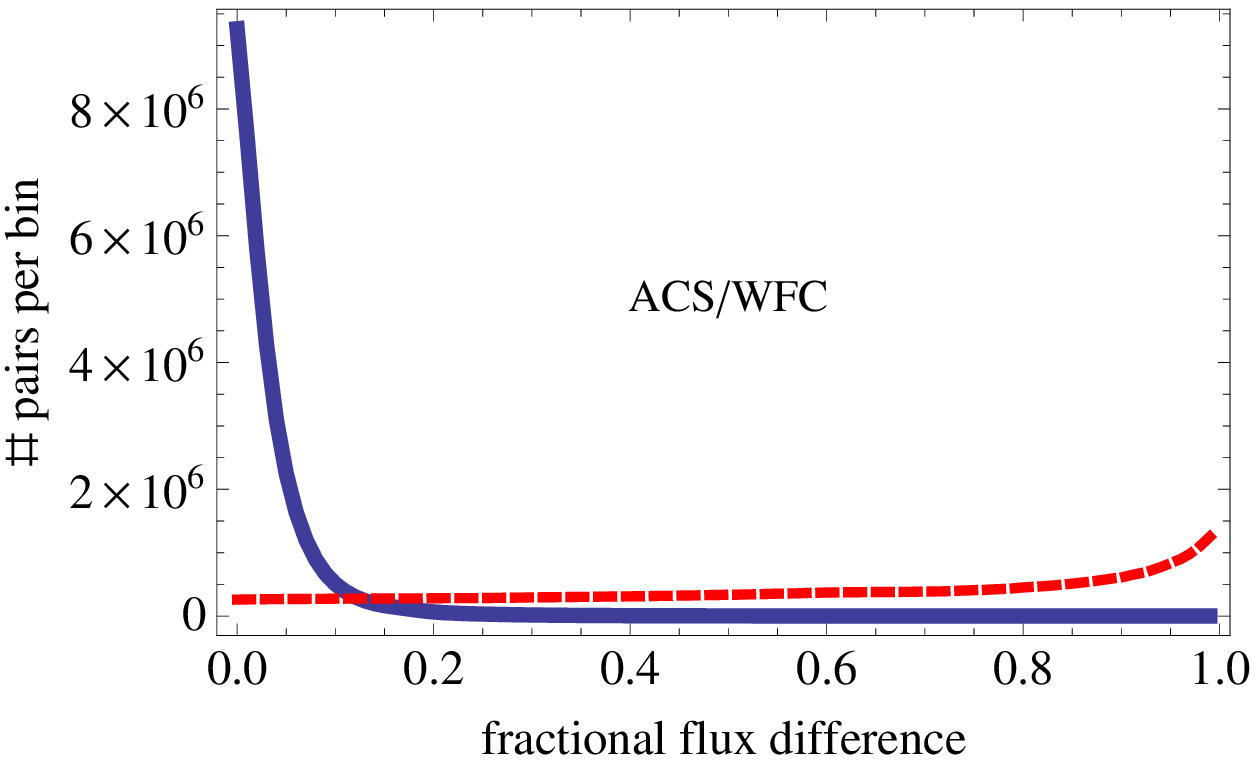}{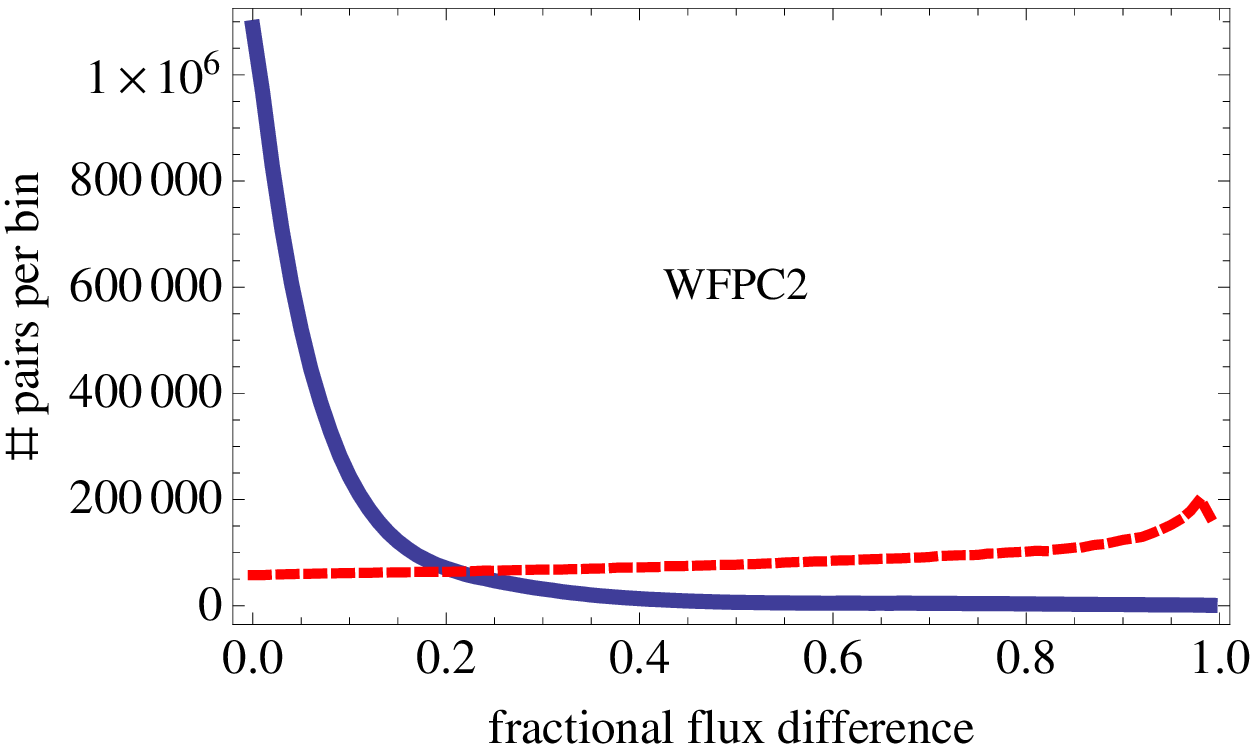}
\caption{
The solid lines plot the distribution of the fractional flux differences between pairs of sources (defined
in Equation (\ref{fracfl})) that are in the same match
and have the same instrument, detector and filter. The bin size is 0.01.
The left panel is for ACS/WFC and the right panel is for WFPC2.
Plotted as dashed lines are the distributions with 
pairs selected randomly from the same images as each of the pairs used for the plots in the solid lines. }
\label{fig:dfluxmix}
\end{figure}

We define the fractional flux difference between
source pairs $i$ and $j$ having fluxes $F_i$ and $F_j$ to be
\begin{equation}
f_{ij} = \frac{|F_i - F_j|}{\max(F_i,F_j)}.
\label{fracfl}
\end{equation}
With this definition, we have that $0 \le  f_{ij}  < 1.$
In Fig.~\ref{fig:dfluxmix}, we plot in solid lines the distributions of $f$ for ACS/WFC and WFC2 for pairs
in the same match that have the same instrument, detector and filter.  We compare these results to corresponding cases where pairs are not
matched.
To carry out this comparison, we considered the 
same set of pairs as those used in the respective solid lines and mapped the sources
in each pair to a random pair taken from the same pair of images. 
The figures show that the matched pair distributions have strong peaks near \mbox{$f\!=\!0$}, while the randomly
selected pairs have a very broad distribution with a mild peak near \mbox{$f\!=\!1$}.
The results provide evidence that these matches contain repeated observations of the same physical
object.

\begin{figure}
\epsscale{0.5}
\plotone{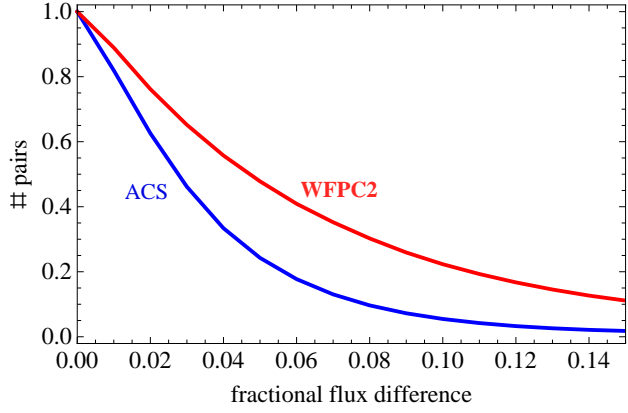}
\caption{
Plot of the distributions of the fractional flux differences between pairs of sources (defined
in Equation (\ref{fracfl})) that are in the same match
and have the same instrument, detector and filter. The bin size is 0.01. The vertical scale is normalized to unity for zero flux difference. }
\label{fig:dfluxacswfpc2}
\end{figure}

A more detailed view of the flux differences for matched pairs is shown in Fig.~\ref{fig:dfluxacswfpc2}.
The width of the ACS flux distribution, based on where normalized distribution equals 0.5, is equal to 0.027 and
the corresponding width of the WFPC2 flux distribution is equal to 0.047. The smallness of the widths 
reflect  the accuracies  of both the matching and the HST photometry. 
Since no constraints
are applied to the exposure times of the matched pairs, some of the larger flux differences likely involve detections
of faint sources with short exposure times. In some other cases, the flux differences may reflect physical
changes in the source brightness.

\begin{figure}
\epsscale{1}
\plottwo{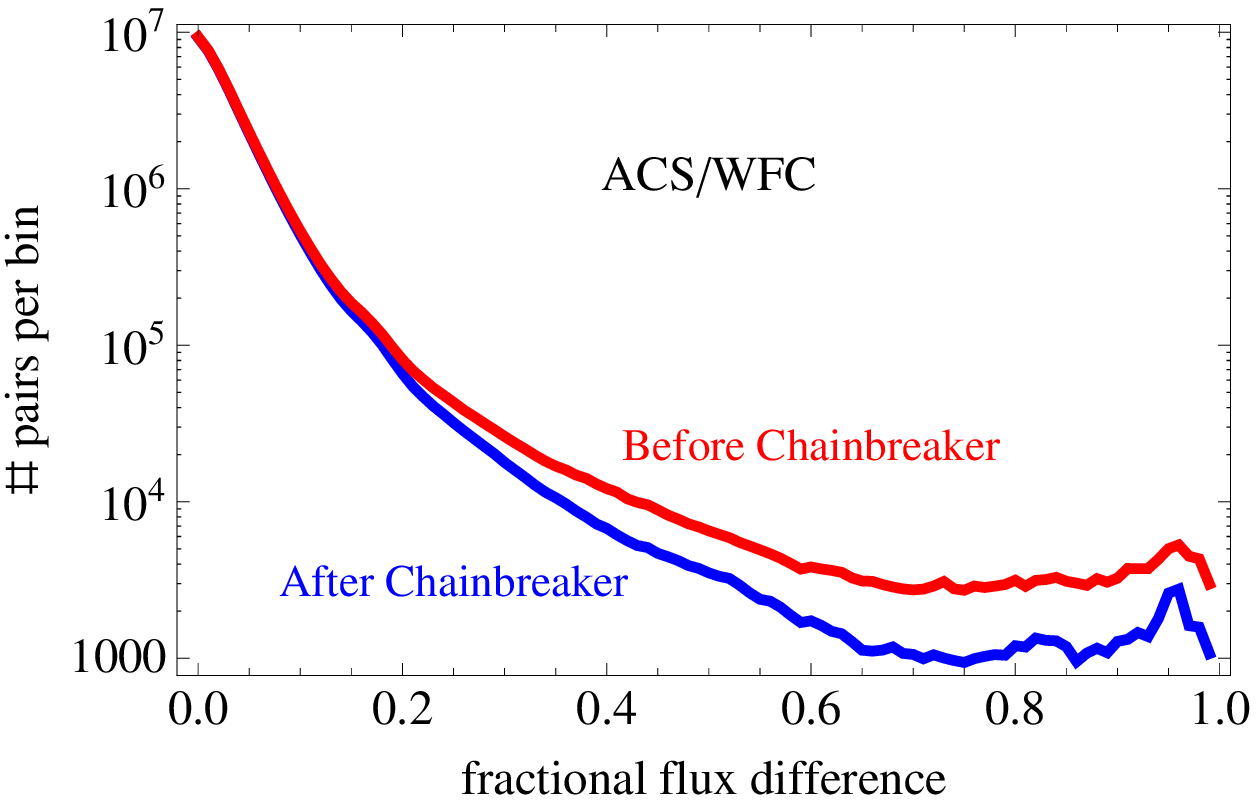}{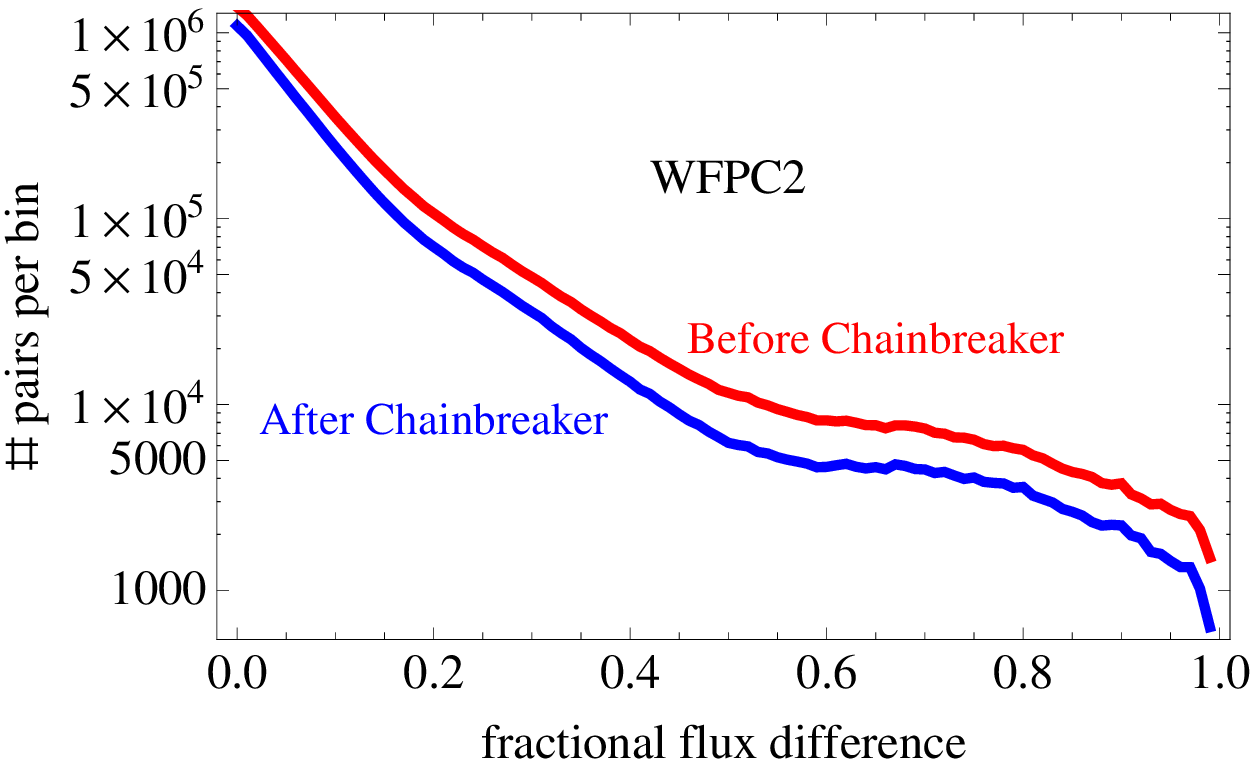}
\caption{
Semi-log plot of the distribution of the fractional flux differences between pairs of sources (defined
in Equation (\ref{fracfl})) that are in the same match
and have the same instrument, detector and filter. The bin size is 0.01. The upper (lower) lines are the results 
before (after) the application of chainbreaker tool that splits chains of matches that are
loosely connected. The lower curves are the same as those of Fig.~\ref{fig:dfluxmix},
but plotted on  a semi-log scale. }
\label{fig:dfluxchain}
\end{figure}

We determine effectiveness of the chainbreaker in splitting unphysical matches
by again considering fractional flux differences between pairs of matching sources that have the
same instrument, detector and filter.  Fig.~\ref{fig:dfluxchain} shows the comparison of the distributions
before and after the application of the chainbreaker. For the case of ACS/WFC, we see
that the chainbreaker was effective in reducing the incidence of unphysical matches
in the tail of the distribution.
We note that the plot is semi-logarithmic and the apparent offset between the two curves involves only a small fraction of the matches.
A smaller improvement was made for the case of WFPC2.

\begin{figure}
\epsscale{1}
\plottwo{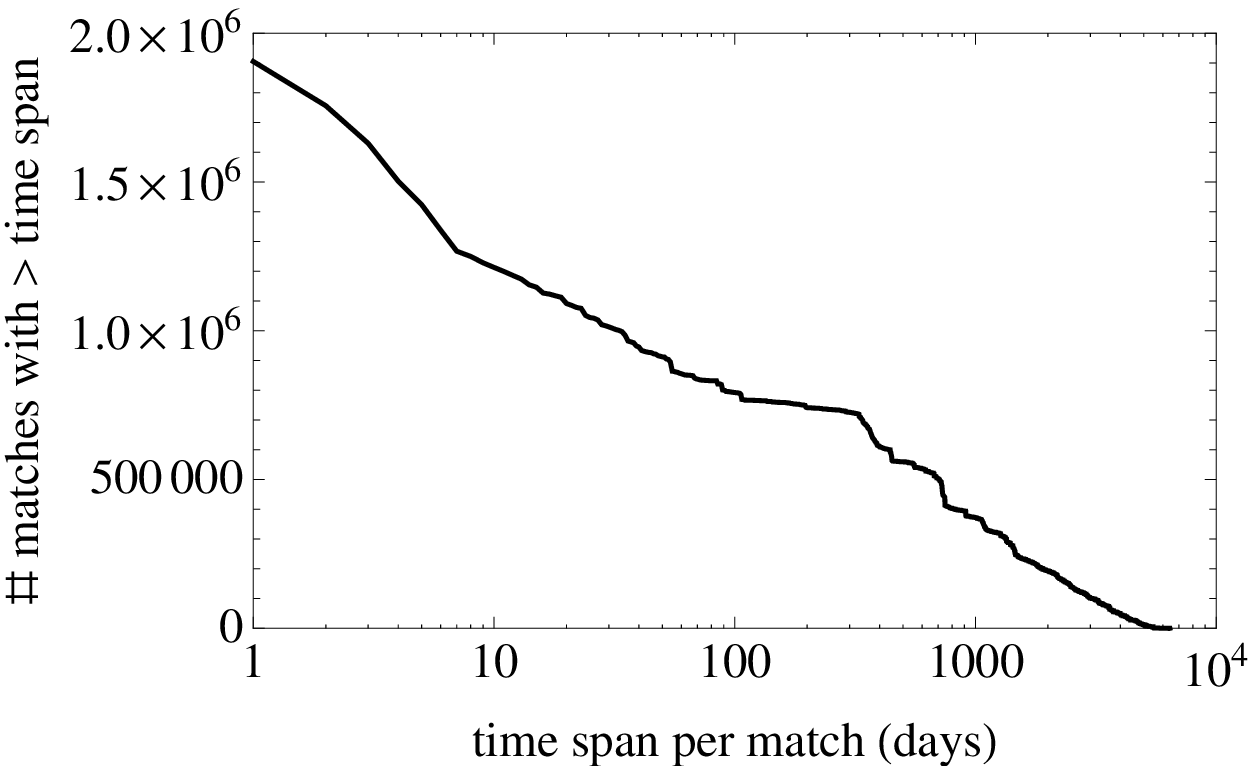}{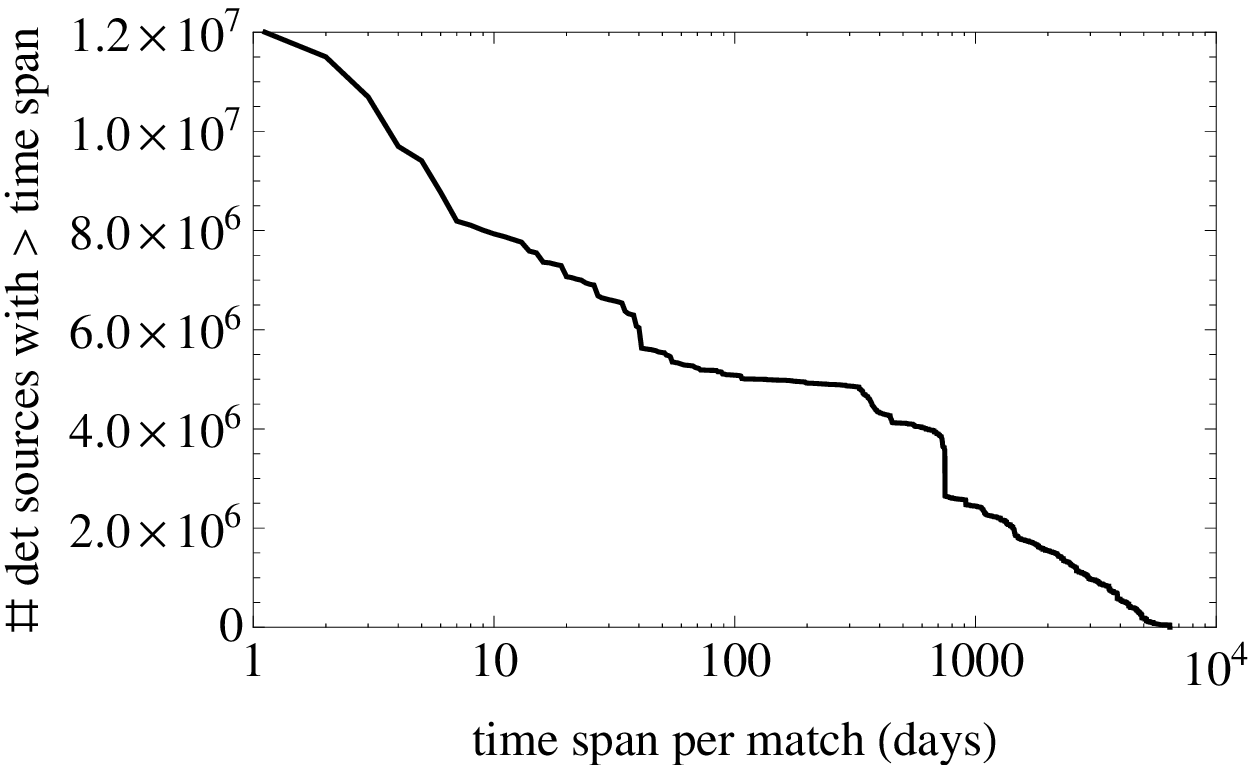}
\caption{
Left: Semi-log plot  of the number of matches that have a time span greater than some value.
Right: Semi-log plot  of the number of sources in matches that have a time span greater than some value. }
\label{fig:timeg}
\end{figure}

We consider the time distribution covered by the matches.
This time coverage of a match
is defined as the difference between the earliest start time and latest stop time
for the exposures containing detected sources in the match. The largest match
time span is 6419 days or about 17.5 years.
The time span distributions are plotted in Fig.~\ref{fig:timeg}.
The number of matches greater than some time span falls off roughly logarithmically
with the time span. Over 1 million matches involving about 10 million source
detections have a time span greater than a few days.
About $7 \times 10^5$ matches involving about 5 million source
detections have a time span greater than a year.
About 10\% of the detected sources lie in matches with a duration of more than a year.
%Among matches with more than one detected source, 7.7\% of the matches involving 17\% of the source detections have a duration of more than one year.
About 30\% of the crossmatched source detections lie in matches that have a duration of more than one year.

\begin{figure}
\epsscale{0.5}
\plotone{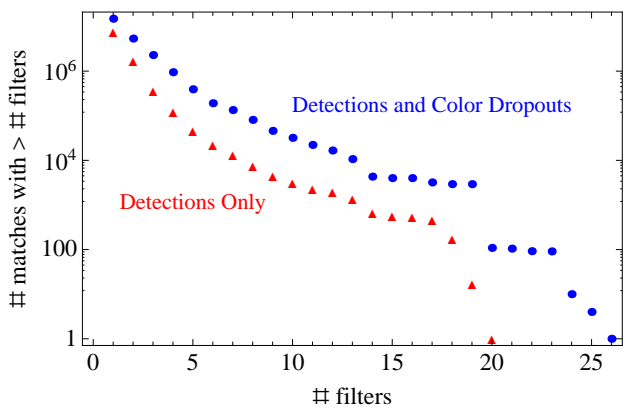}
\caption{
Semi-log plot  of the number of matches that have more filters than some value.
The triangles refer to the number of filters for detected objects. The 
circles refer to refer to the number of filters for detections and color dropouts.
 }
\label{fig:filterg}
\end{figure}

\begin{figure}
\begin{center}
\plotone{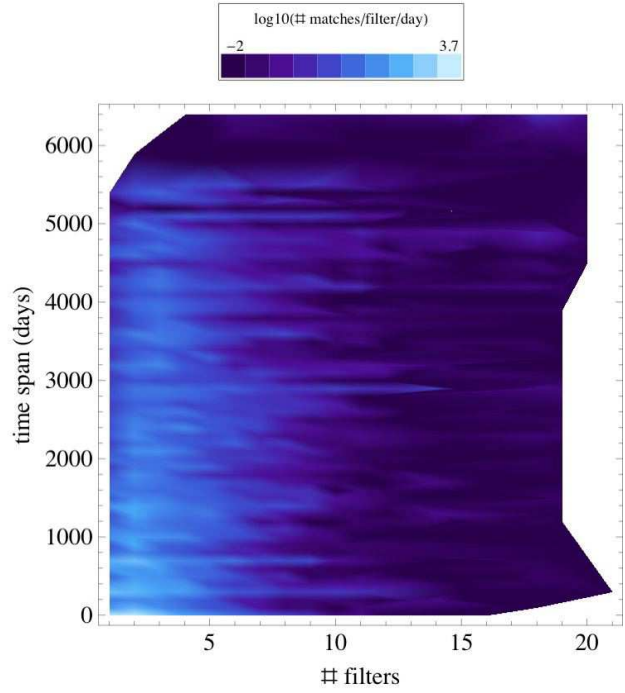}
\end{center}
\caption{Plot of the frequency of matches as a function of the number
of filters in a match and the duration of a match for matches with duration greater than 1 day. 
Only detected sources are considered.
The frequency of matches is color coded
 and taken as the $\log_{10}$ of the number of matches in a bin of 1 filter $\times$ 1 day.   }
\label{fig:filtertimedens}
\end{figure}

Many of the matches involve more than one filter.
Fig.~\ref{fig:filterg}  shows that more than one million matches involve
multiple filters. Matches involve as many as 21 filters of detected sources
and as many as 27 filters of detected sources and color dropouts.
More than $7\times 10^4$ matches involve more than 5 filters for detected sources
and more than $3.9 \times 10^5$ matches involve more than 5 filters for detected sources
and dropouts.
Fig.~\ref{fig:filtertimedens} shows how the frequency of matches depends
jointly on the number of filters and the time duration of a match.
The minimum match duration considered is one day which implies that all
matches being considered involve more than HST visit. The frequency
of matches extends to long time spans for matches with less than 5 filters.
There are bands of many-filter cases at several time spans.

\section{Summary} \label{sec:summary}
\noindent
We developed a new approach for positional cross-matching astronomical sources that is well suited to the Hubble Space Telescope (HST).  We applied the approach to crossmatching HST sources detected
in the same or different detectors (ACS/WFC and WFPC2) and filters.

The HST observations comprise a unique astronomy resource. Preserving the measurements and enhancing their value are important but challenging tasks. While the overall volume of data is moderate by today's standards, the dataset presents a number of difficulties.
One of them is the small field of view that does not contain enough calibrators to accurately pin down the astrometry of the images. We presented a new algorithm that can cross-calibrate overlapping images to each other. We introduced infinitesimal 3-D rotations for this purpose, which yield an analytically tractable optimization procedure.
Our implementation of this scheme is sufficient to provide high-precision relative astrometry across overlapping images. The improved astrometric accuracy of source positions is typically about a few milli-arcseconds (see Fig.~\ref{fig:sigmaC}). 
Another challenge is to deal with the complex placement of the images on the sky and the fact that certain parts of the sky are observed many times (see Fig.~\ref{fig:mossize}). Instead of the naive combinatorial scaling, we achieve high efficiency in a  greedy ``chainbreaker'' procedure that applies Bayesian model selection to find the best matches within  a mosaic. 

To check on our matching results,
we analyzed the flux differences between sources in the same match with the same instrument, 
detector and filter, which should  optimally be zero (ignoring variable sources).  We found that the astrometric correction and the probabilistic object selection provide reliable matches  (see Fig.~\ref{fig:dfluxmix}). Based on just positional information, the Bayesian model selection rejects spurious matches and improves the tail of the distribution with large flux deviations (see Fig.~\ref{fig:dfluxchain}).

We demonstrated that many of the matches cover a broad range of time spans and filters
(Figs.~\ref{fig:timeg}, \ref{fig:filterg}, and \ref{fig:filtertimedens}). Therefore, the catalog
should enable time-domain, multi-wavelength studies of sources detected by HST. The catalog
also provides information about nondetections.
The presented catalog is publicly available online\footnote{The catalog is available at \url{http://archive.stsci.edu/hst/hla\_cat/}} via Web forms
 as well as an advanced query interface. 

The catalog provides a basis for further extensions, such as by including
other  detectors and source lists based on other software.
The algorithms and tools developed in this paper are not specific to the Hubble Space Telescope. They are directly applicable to any astronomy observations that exhibit similar challenges.

\section*{Acknowledgements} \noindent
%\acknowledgements
We benefited from inspiring discussions with Alex Szalay, Rick White, and Brad Whitmore on different aspects of the project.
We acknowledge assistance on a previous approach to this project  by Nathan Cole.
We are grateful for support from NASA AISRP grant NNX09AK62G.

\appendix
\section*{APPENDIX}
\section{Derivation of the Astrometric Correction\label{app1}}
\noindent
In this section we provide a short derivation for Equation~(\ref{omeq}) using a variational method. 
This approach to the minimization is equivalent to writing out the components of the vectors and matrices to obtain the vanishing partial differentials, but more concise.
The quadratic cost function is
\begin{equation}
C(\Vom)= \sum_{\bi} \wb \Big[\Vcb - \big(\Vrb + \Vom\times\Vrb\big) \Big]^2 \ .
\end{equation}
Its minimization yields the value of $\VomHat$ defined by Equation~(\ref{vomhat}), that is, 
\begin{equation}
\VomHat = \argmin_{\Vom \in \mathbb{R}^3} C(\Vom) \ .
\end{equation}
To determine the $\VomHat$ solution, we consider small variations in $C(\Vom)$ for \mbox{$\Vom\simeq\VomHat$} and require that they vanish. This procedure is equivalent to requiring that
the partial derivatives of $C(\Vom)$ with respect to the components of $\Vom$ vanish at \mbox{$\Vom=\VomHat$}. 
The linear variation of the cost function due to a small $\delta \Vom$ change is %in $\Vom$ is
\begin{equation}
\delta C(\Vom)= 
	-2 \sum_{\bi} \wb \Big[\Vcb - \big(\Vrb + \Vom\times\Vrb\big) \Big] \cdot 
	(\delta \Vom\times\Vrb) \ .
\label{dC}
\end{equation}
We use the fact that the terms in a scalar triple product can be cyclically permuted as in
\begin{equation}
  (\Vcb -\Vrb) \cdot (\delta \Vom\times\Vrb) =  \delta \Vom \cdot \big[\Vrb \times(\Vcb -\Vrb)\big] \ ,
\end{equation}
and the known equivalence for the vector triple products, in addition to that for the dot products of two cross products,
\begin{eqnarray}
 (\Vom\times\Vrb) \cdot (\delta \Vom\times\Vrb)  &=& \delta \Vom \cdot 
 \big[\Vrb \times  (\Vom\times\Vrb) \big]\\
   &=& -\delta \Vom \cdot \left[ \Vrb\,(\Vrb \cdot 
   \Vom) - r_i^2 \, \Vom   \right] \ .
\end{eqnarray}
Since all $\Vrb$ are unit vectors, i.e., \mbox{$r_i\!=\!1$}, Equation~(\ref{dC}) can now be written as
\begin{equation}
\delta C(\Vom)= 
	-2 \, \delta \Vom \cdot  
	\sum_{\bi} \wb \Big[ \Vrb \times(\Vcb -\Vrb) +  \Vrb\,(\Vrb \cdot \Vom) - \Vom  \Big] \ .
\end{equation}
By requiring that \mbox{$\delta C(\VomHat)=0$} for arbitrary, but small $\delta \Vom$, we have that
\begin{equation}
\sum_{\bi} \wb \Big[ \Vrb \times(\Vcb -\Vrb) +  \Vrb\,(\Vrb \cdot \VomHat) - \VomHat  \Big]  = 0 \ .
\end{equation}
The cross-product of any vector by itself vanishes, so 
\mbox{$\Vrb\!\times\!(\Vcb\!-\!\Vrb) = \Vrb\!\times\!\Vcb$}, and
the final result is
\begin{equation}
\sum_{\bi} \wb \Big[ (\Vrb\times\Vcb) + \big(\Vrb\otimes\Vrb-\mI\big)\, \VomHat  \Big]  = 0 \ .
\end{equation}
Symbol $\mI$ represents the identity and $\otimes$ is the dyadic vector product, hence the term in parenthesis is a linear operator applied to the $\VomHat$ vector.
This formula is equivalent to Equation~(\ref{omeq}) in the main body of the article.

\newpage

\end{document}